\documentclass[journal,draftclsnofoot,onecolumn,twoside,12pt]{IEEEtran}

\usepackage[T1]{fontenc}
\usepackage{amsmath} 
\usepackage{amssymb}
\usepackage{algorithmic}
\usepackage{cite}
\usepackage{dsfont} 
\ifCLASSINFOpdf
   \usepackage[pdftex]{graphicx}
\else
  \usepackage[dvips]{graphicx}
\fi

\newtheorem{lemma}{Lemma}
\newtheorem{claim}{Claim}
\newtheorem{remark}{Remark}

\DeclareMathOperator{\Tr}{Tr}
\DeclareMathOperator{\E}{E}

\DeclareMathOperator*{\argmax}{argmax}
\DeclareMathOperator{\diag}{diag}
\DeclareMathOperator{\rank}{rank}

\begin{document}
\title{User Selection and Widely Linear Multiuser Precoding for
One-dimensional Signalling}

\author{Majid Bavand, \IEEEmembership{Student Member, IEEE,} Steven D. Blostein,
\IEEEmembership{Senior Member, IEEE}
\thanks{
The authors are with the Department
of Electrical and Computer Engineering, Queen's University, Ontario,
Canada, K7L 3N6 (e-mail: \{m.bavand, steven.blostein\}@queensu.ca).}
}
\maketitle

\begin{abstract}
Massive deployment of low data rate Internet
of things and ehealth devices
prompts us to develop more practical precoding and user selection techniques
that comply with these requirements.
Moreover, it is known that when the data is real-valued and the observation is complex-valued,
widely linear (WL) estimation can be employed in lieu of linear estimation to improve
the performance. With these motivations,
in this paper, we study the transmit precoding (beamforming)
in multiuser multiple-input single-output
communications systems assuming the transmit signal is one-dimensionally modulated
and widely linear estimation is performed at the receivers.
Closed-form solutions for widely linear maximum ratio transmission (MRT), WL zero-forcing (ZF),
WL minimum mean square error (MMSE), and WL maximum signal to leakage and noise ratio (MSLNR)
precoding are obtained. It is shown that widely linear processing can potentially
double the number of simultaneous users compared to the linear
processing of one-dimensionally modulated signals.
Furthermore, to deal with the increasing number of communications devices
a user selection algorithm compatible with widely linear processing of
one-dimensionally modulated signals is proposed.
The proposed user selection algorithm
can double the number of simultaneously selected users compared to conventional
user selection methods.
\end{abstract}

\section{Introduction}\label{secIntro_wl}
\IEEEPARstart{W}{idely} linear (WL) processing of complex-valued signals,
originally introduced in \cite{Crane69} and later
resurrected in \cite{Chevalier95} in the context of minimum mean square
error estimation, refers to the superposition of linear filtering of the observation
and linear filtering of its complex conjugate, or equivalently,
superposition of linear filtering
of real and imaginary parts \cite{Utschick13}. The latter representation
is known as the composite real representation. It was shown
in \cite{Chevalier95} that when the distribution of the estimand (signal of
interest) is improper, i.e., not circularly symmetric, widely
linear estimation will improve  mean square error (MSE)
estimation, whether or not the observation is improper.
Since its resurrection by Picinbono and Chevalier \cite{Chevalier95}, WL
processing has been applied to communications systems,
specifically to improper signal constellations
or when improper noise is encountered \cite{Utschick13,MajidGc16}.

Advances in wireless communications, in conjunction with advances
in electronics, paved the way for emergence of
technologies including the Internet of things (IoT)
and pervasive ubiquitous ehealth
\cite{Perera14,Nokia15,Pantelopoulos10,Patel10}.
In wireless communications,
low-data-rate power efficient one-dimensionally (1D) modulated
signals such as binary phase shift keying (BPSK) are of
interest to reliably support these emerging systems with massive numbers
of low-data-rate devices \cite{Schwarz16,Nokia15}.
When the data is one-dimensionally modulated,
implying an improper distribution, widely
linear estimation has been applied to receive beamforming
in the context of multiple-antenna communications
\cite{Schreier03,Chevalier06,Chevalier09,Hanzo06}.
In contrast to linear receive beamforming, in which the
output is given by a linear spatial filter ${\bf w}$ applied to received
signal ${\bf r}$ as $y = {\bf w} {\bf r}$, a widely linear receive beamformer
output is given by $y = {\bf w}{\bf r} + {\bf v}{\bf r}^*$, which includes superposition
of linear beamforming ${\bf v}$ of the complex
conjugate of the received signal. For the
special case of one-dimensionally modulated signals, widely
linear receive beamforming reduces to the real part of the
linearly filtered observation, i.e., $y = \Re\{{\bf w}{\bf r}\}$ \cite{Chevalier95}.


The concept of widely linear processing can also be applied
to transmit precoding (beamforming). From one
perspective, widely linear precoding is the superposition of
linear precoding of the modulated signal and linear precoding
of the complex conjugate of the modulated signal \cite{ZhangTSP13}.
From another perspective, widely linear precoding is the linear
precoding of the modulated signal in conjunction with
widely linear estimation of the received signal \cite{Sellathurai10}. 
It should be remarked that if the modulated signal is real-valued,
only the latter perspective of widely linear transmit
precoding is relevant.

In wireless systems, the base station or access point
may be equipped with multiple antennas and users are typically
equipped with a single antenna due to physical constraints
such as equipment size, power supply, cost, and computational
capabilities \cite{Leung13}. Consequently, the downlink transmitter can
transmit different data streams to multiple users simultaneously
to exploit the available spatial multiplexing gain.
The task of transmit precoding is to reduce the effect of co-channel
interference which arises in wireless broadcast channels due
to spatial multiplexing. Four basic linear transmit precoders
that are well researched in the last decades are: (i) transmit
matched filtering or maximum ratio transmission (MRT) precoding
which maximizes the signal portion of the desired signal
at each receiver \cite{Nossek05}, (ii) transmit zero-forcing or channel
inversion precoding which nulls the interference at each receiver
\cite{Peel05,Nossek05}, (iii) transmit minimum mean square error
or regularized channel inversion precoding which minimizes
sum of mean square errors (MSE) of users \cite{Yener04,Peel05,Nossek05}, and
(iv) maximum signal to leakage and noise ratio (MSLNR)
precoding \cite{Sadek07WC}. In this paper, we develop widely linear counterparts
of these four basic linear precoding techniques for
one-dimensional signalling.

As mentioned earlier, emerging applications such as IoT and ehealth traffic
require support of a large number of low-data-rate users.
To support large numbers of devices, one of the challenging issues is network traffic.
One approach to reduce the network traffic is to increase spectral efficiency by
simultaneously transmitting information to multiple users using transmit precoding
techniques. User selection is another approach that can be combined with transmit precoding
to improve spectral efficiency \cite{Goldsmith06}.
The choice of the best user subset, which depends on the precoding
method, is critical in this scenario.
Existing low-complexity user selection techniques that account for the
interference arising in multiuser communications in broadcast channels are
capable of selecting, at most, as many users as the number of
transmit antennas \cite{Evans06,Leung13,Goldsmith06}.
In \cite{MajidArXiV16}, it has been shown that geometric user selection (GUS)
algorithm is capable of overloading the system with more
simultaneous users than the number of transmit antennas,
in a system with one-dimensionally modulated signals and
minimum probability of error (MPE) precoding.
As has been shown in \cite{MajidArXiV16}, although the computational complexity of GUS
algorithm is very low, it does not always select more users
compared to other existing user selection methods, especially
when the number of available users is not large enough.
In this paper, inspired by the semi-orthogonal user selection
(SUS) algorithm \cite{Goldsmith06}, a semi-orthogonal user selection method
for one-dimensional modulation (SUSOM) is developed. It is
shown that SUSOM is able to double the number of selected
users. In other words, a transmitter with $M$ transmit antennas
is shown to be capable of supporting at most $2M$ simultaneous
users.

The rest of the paper is organized as follows:
Section \ref{secSysModel_wl} introduces our system model.
Section \ref{secWidelyLinear_wl},
studies the WL MRT, WL ZF, WL MMSE, and WL MSLNR precoding for
one-dimensional signalling. Closed-form solutions for the precoders of the WL MRT
and the WL ZF are obtained by using complex-domain analysis and
closed-form solutions of the WL MMSE and the WL MSLNR precoders are obtained by
analysis of the composite real representation.
Section \ref{secSusom_downlink} introduces semi-orthogonal
user selection for one-dimensional modulation with capability of overloading
the system with more users than the number of transmit antennas.
Numerical results are presented in Section \ref{secResults_wl}.
Finally, conclusions are drawn in Section \ref{secConclusion_wl}.


\section{System Model}\label{secSysModel_wl}
A multiuser multiple-input single-output wireless broadcast channel
with an $M$-antenna transmitter and $K$ single-antenna users is considered.
The transmitter is assumed to simultaneously send independent pulse amplitude modulated (PAM)
signals to all users using the same carrier frequency and bandwidth.
In low-pass vector space representation,
the one-dimensionally modulated signal of a user can be described by
a real-valued scalar which is the projection of the low-pass representation of the signal over
the basis function defined as $f(t) = {g(t)}/{\sqrt{E_g}}$,
where $g(t)$ is the low-pass real-valued pulse shaping signal in the interval $0 \le t \le T$,
with power $E_g = \frac{1}{2T}\int_{0}^{T} {g^2(t) dt}$. Therefore, the PAM
signal of user $k$ can be represented by
\begin{equation}\label{pamSet_wl_eq}
    s_k(l_k) \in \{(2l_k-1-L_k)d \sqrt{E_g}| ~1 \le l_k \le L_k \}, \quad 1 \le k \le K.
\end{equation}
The modulation order of user $k$ is denoted by $L_k$, i.e.,
the total number of constellation points in the pulse amplitude modulated signal of user $k$
is $L_k$. This also implies that different users may not necessarily employ the same
modulation order.
The distance between adjacent signal constellation points is $2d\sqrt{E_g}$.
Given $l_k$, the power of the signal is $s_k^2(l_k)$.
Consequently, the average power of the modulated signal of user $k$
is $\sigma^2_{s_k} = \frac{L_k^2 - 1}{3}d^2 E_g$.
Using an $M\times 1$ precoding vector ${\bf u}_k$
to encode the symbol transmitted to user $k$,
the transmitted signal is then given by
\begin{equation}\label{precodedSignal_wl_eq}
    {\bf x} = \sum_{k=1}^K {\bf u}_k s_k = {\bf U} {\bf s},
\end{equation}
where ${\bf U} = \left[{\bf u}_1, \cdots, {\bf u}_K\right]$,
${\bf s} = [s_1, \cdots, s_K]^T$, and $s_k = s_k(l_k)$.
Therefore, the transmit power is expressed by
\begin{equation}\label{txPower_wl_eq}
    \E[\|{\bf x}\|^2_2] = \Tr({\bf U}{\bf R}_s{\bf U}^H),
\end{equation}
where it is assumed that the input signals are mutually independent
with covariance matrix ${\bf R}_s = \E[{\bf s}{\bf s}^T] =
{\rm diag}(\sigma_{s_1}^2,\ldots,\sigma_{s_K}^2)$.

Assuming a fading channel with additive white Gaussian noise (AWGN), the received
signal $r_k$ at user $k$ is given by
\begin{equation}\label{rxSignal_wl_eq}
    r_k = {\bf h}_k{\bf x} + z_k, \quad 1 \le k \le K,
\end{equation}
where the additive noise $z_k$ is a circularly
symmetric complex Gaussian (CSCG) random variable with
zero mean and variance $\sigma_{z_k}^2$, and
the $1 \times M$ vector ${\bf h}_k$ is the channel between the $M$ antennas of
the transmitter and the single antenna of user $k$.
The entries of ${\bf h}_k$ follow an independent identically distributed (i.i.d.)
CSCG distribution with zero mean and variance 1.
This channel model is valid for narrowband (frequency non-selective) systems if the
transmit and receive antennas are in
non line-of-sight rich-scattering environments with sufficient antenna spacing
\cite{PaulrajBook03,Mao12}.
Equivalently, (\ref{rxSignal_wl_eq}) can be represented in vector form by
\begin{equation}\label{rxSignalVector_wl_eq}
    {\bf r} = {\bf H} {\bf x} + {\bf z},
\end{equation}
where ${\bf r} = [r_1,\ldots,r_K]^T$, ${\bf H} = [{\bf h}^T_1,\ldots,{\bf h}^T_K]^T$,
and the noise ${\bf z} = [z_1,\ldots,z_K]^T$ has zero mean and covariance matrix
${\bf R}_z = \diag(\sigma_{z_1}^2, \cdots, \sigma_{z_K}^2)$.

The received signal at each user is passed through a filter. Therefore, the
processed signal at the receiver of user $k$ is represented as a function of the
transmit precoding matrix ${\bf U}$ and the receive filtering
coefficient $w_k$ by
\begin{align}\label{processedSignal_wl_eq}
    \nonumber
    y_k &= w_k r_k =
    w_k {\bf h}_k {\bf U} {\bf s} + w_k z_k =
    \sum_{j=1}^K w_k {\bf h}_k {\bf u}_j s_j + z'_k \\
    &= w_k {\bf h}_k {\bf u}_k s_k + w_k {\bf h}_k {\bf U}_{\bar k} {\bf s}_{\bar k} + z'_k
    , \quad 1 \le k \le K,
\end{align}
where $z'_k$ is also a CSCG noise term\footnote{Affine transformation
preserves properness (circular symmetry) of a random variable \cite{Massey93}.} with
variance $\sigma_{z'_k}^2=\sigma_{z_k}^2 w_k w_k^*$,
${\bf s}_{\bar k} = [s_1, \cdots, \allowbreak s_{k-1}, \allowbreak s_{k+1}, \cdots, s_K]^T$, and
${\bf U}_{\bar k} = [{\bf u}_1, \cdots, {\bf u}_{k-1}, {\bf u}_{k+1}, \cdots, {\bf u}_K]$.
Equivalently, the processed signals at the receivers can be represented in vector form by
\begin{equation}
    {\bf y} = {\bf W}{\bf H}{\bf U} {\bf s} + {\bf z'},
\end{equation}
where ${\bf W} = \diag(w_1, \ldots, w_K)$ and ${\bf y} = [y_1,\ldots,y_K]^T$,
and ${\bf z'} = {\bf W} {\bf z}$ and has zero mean and covariance matrix
${\bf R}_{z'} = {\bf W} {\bf R}_z {\bf W}^H$.

Since the focus of this paper is on precoding design, it is assumed
that ${\bf W}$ is known at the transmitter. When the structure
of the receivers are required to be simple without any filtering, received
filtering coefficients can be assumed to be equal to 1.
Moreover, it is assumed that perfect channel state information
between transmitter and all users is available at
the transmitter in order to focus on the precoding methods rather than on the effect
of channel estimation. This information could be obtained, for example, by using feedback
and pilot-based estimation at the receivers or by assuming time division duplex (TDD)
systems.

\section{Widely Linear (WL) Processing}\label{secWidelyLinear_wl}
Widely linear (WL) processing was resurrected by Picinbono and Chevalier in the context
of mean square error estimation of complex-valued data \cite{Chevalier95}.
In general, when data
$s$ and observation $r$ are both improper and complex, widely linear
estimation of $s$ is given by ${\hat s} = w r + v r^*$, i.e.,
by superposition of linear estimates of observation $r$ and its complex
conjugate $r^*$. In case of real-valued data and complex-valued observation,
it is known that $v =  w^*$ and therefore ${\hat s} = 2 \Re\{w r\}$,
i.e., the estimation is given by the real part of the output of a linear
estimator \cite{Chevalier95}.
It is expected that, calculating $w$ by optimizing a metric based on $\Re\{w r\}$
provides more degrees of freedom compared to optimizing a metric
based on $w r$. In other words, by not using the information hidden in
the imaginary part of the output, $\Im\{w r\}$, WL processing is expected to be capable of
providing more degrees of freedom (DoF) which could be utilized either for
improving reliability or throughput.

Now let us consider the system model introduced in Section \ref{secSysModel_wl}.
Using the midpoints between the received signal constellation points as
the decision thresholds \cite{ProakisBook01}, the following widely linear decision rule
may be used for estimating the transmitted PAM symbols of user $k, ~ 1 \le k \le K$:
\begin{align}\label{pamDecision_wl_eq}
    \nonumber
    &{\hat s}_k \\
    &= \left\{ \begin{array}{lr}
    s_k(1) & y_k^R \!\le\! \Re\{w_k {\bf h}_k {\bf u}_k s_k(1) \!+\!
    w_k {\bf h}_k {\bf u}_k d \sqrt{E_g}\}  \\[.1in]
    s_k(l_k) & \begin{array}{r} \Re\{w_k {\bf h}_k {\bf u}_k s_k(l_k) \!-\!
    w_k {\bf h}_k {\bf u}_k d \sqrt{E_g}\} \\
    < y_k^R \le \\
    \Re\{w_k {\bf h}_k {\bf u}_k s_k(l_k) \!+\! w_k {\bf h}_k {\bf u}_k d \sqrt{E_g}\};\\
    2 \le l_k \le L_k-1 \end{array} \\[.35in]
    s_k(L_k) & y_k^R \!>\! \Re\{w_k {\bf h}_k {\bf u}_k s_k(L_k) \!-\!
    w_k {\bf h}_k {\bf u}_k d \sqrt{E_g}\}
    \end{array} \right.\!\!\!,
\end{align}
where the superscript $^R$ denotes the real part, i.e., $x^R={\Re}\{x\}$.
Therefore, since $y_k^R$ is considered in calculating the receive beamformer or
the transmit precoder rather than $y_k$, the processing is considered as being widely linear
rather than linear.

\subsection{WL Maximum Ratio Transmission Precoding}\label{subSecMrt_wl}
Maximum ratio transmission (MRT) or matched filtering is the transmit
counterpart of maximum ratio
combining at the receiver \cite{Nossek05}. MRT intends to maximize the received signal to noise
ratio by matching the transmit precoding vector of each user to its channel. Since
MRT does not consider co-channel interference, it may only perform close to optimally
in noise-limited channels and in single-user communications. Considering widely linear processing,
the SNR at the $k$th user, $1 \le k \le K$, can be defined as the ratio of the power
of the real part of the desired signal at receiver $k$ to the power of the real part
of the post-processing noise, i.e.,
\begin{equation}
    \text{SNR}_k =
    \frac{\E[|\Re\{w_k {\bf h}_k {\bf u}_k s_k\}|^2]}{\E[|\Re\{z_k^\prime\}|^2]}
    = \frac{2\sigma_{s_k}^2 (\Re\{ w_k {\bf h}_k {\bf u}_k \})^2 }{\sigma_{z'_k}^2}.
\end{equation}
Then the MRT precoding problem can be formulated by maximizing the SNR at the receiver
subject to a constraint on the transmit power as
\begin{subequations}\label{mrtBfProblem_wl_eq}
\begin{align}
    &\max_{{\bf u}_k } \text{SNR}_k \\
    &{\text{subject to}} \quad \sigma_{s_k}^2 \|{\bf u}_k\|_2^2 \le \tau_k,
\end{align}
\end{subequations}
for $1 \le k \le K$. In (\ref{mrtBfProblem_wl_eq}), $\tau_k$, the power
constraint on the transmitted signal to user $k$, should
satisfy $\sum_{k=1}^K \tau_k = \tau$, where $\tau$ is the total transmit
power constraint. 
To calculate the values of $\tau_k$s, a power allocation strategy such as
equal power allocation or sum rate maximizing water-filling power
allocation can be employed \cite{Goldsmith06}.
Using the Cauchy-Schwarz inequality the solution can be shown to be
\begin{equation}
    {{\bf u}_{k}}_{\text{MRT}} = \frac{\sqrt{\tau_k}}{\sigma_{s_k}}
    \frac{{\bf h}_k^{\prime H}}{|{\bf h}_k^\prime|},
\end{equation}
or in matrix form
\begin{equation}\label{wlMrtMatrix_eq}
    {\bf U}_{\text{MRT}} = {\bf H}^{\prime H} \pmb\Lambda,
\end{equation}
where ${\bf h}'_k = w_k {\bf h}_k$,
${\bf H}' = {\bf W}{\bf H} = [{\bf h}^{\prime T}_1,\ldots,{\bf h}^{\prime T}_K]^T$,
and $\pmb\Lambda = \diag(\frac{\sqrt{\tau_1}}{\sigma_{s_1} |{\bf h}'_1|},\cdots,
\frac{\sqrt{\tau_K}}{\sigma_{s_K} |{\bf h}'_K|})$.
Interestingly, (\ref{wlMrtMatrix_eq}) is the same as the result of MRT with
linear processing in the broadcast channel \cite{Ottersten14}. In other words,
using widely linear processing is not advantageous compared to linear
processing when the transmitter uses MRT to transmit
one-dimensionally modulated signals.

\subsection{WL Zero-Forcing Precoding}\label{subSecZf_wl}
Next, we consider zero-forcing (ZF) precoding also known as
channel inversion \cite{Nossek05}.
In zero-forcing, it is assumed that the received signals are interference free,
i.e., the received signal at receiver $k$ is free of interference caused by the
signal transmitted to user $j \in \{1,\ldots, K\} \setminus \{k\}$. In other words,
zero interference imposes the following constraint on the precoding matrix:
\begin{equation}\label{zfConstraint_wl_eq}
    {\bf W} {\bf H}{\bf U} = {\pmb \Lambda},
\end{equation}
where non-negative real-valued diagonal matrix
${\pmb\Lambda} = \diag(\sqrt{\lambda_1},\cdots,\sqrt{\lambda_K})$. Imposing
this constraint results in
$\sigma_{s_k}^2\lambda_k$ as the average power of the received signal at receiver $k$.
Since ZF precoding only considers the effect of interference but not noise,
it may only perform close to optimally in interference-limited channels.
From (\ref{zfConstraint_wl_eq}) it can be seen that ZF not only forces interference
to be zero, but also it makes the power received by each user to be fixed (and not
necessarily equal).

When the transmitter sends one-dimensionally modulated signals to the users and
estimation of the received signal is performed only over the real part of the
received signal (\ref{pamDecision_wl_eq}), widely linear processing can be employed
which results in relaxing (\ref{zfConstraint_wl_eq}) to
\begin{equation}\label{wlZfConstraint_wl_eq}
    \Re\{{\bf W} {\bf H} {\bf U}\} = {\pmb \Lambda}.
\end{equation}
Thus, as an extension to linear ZF precoding \cite{Nossek05},
widely linear zero-forcing precoding is formulated by minimizing the total
transmit power subject to the interference-free constraint of (\ref{wlZfConstraint_wl_eq})
as
\begin{align}
    \nonumber
    &\min_{\bf U} \Tr({\bf U}{\bf R}_s {\bf U}^H)\\
    &{\text{subject to}} \quad \Re\{{\bf W} {\bf H}{\bf U}\} = {\pmb \Lambda}.
\end{align}

To solve this problem we first rewrite it in the following form:
\begin{align}\label{zfBfModified_wl_eq}
    \nonumber
    &\min_{{\bf u}_1,\ldots,{\bf u}_K}
    \sum_{k=1}^K \sigma_{s_k}^2 {{\bf u}_k^H{\bf u}_k}\\
    &~\!{\text {subject to}} \quad \Re\{{\bf H}'{\bf u}_k\} - \sqrt{\lambda_k}{\bf e}_k = {\bf 0},
    \quad 1 \le k \le K,
\end{align}
where ${\bf e}_k$ is the $k$th standard basis vector in $K$-dimensional Euclidean space
and ${\bf H}'$ was introduced in Section \ref{subSecMrt_wl}.
Accordingly, the Lagrangian is given by
\begin{align}\label{zfLagrangian_wl_eq}
    \nonumber
    &L({\bf u}_1,\ldots,{\bf u}_K,{\pmb\mu}_1,\ldots,{\pmb\mu}_K) \\
    &=
    \sum_{k=1}^K{\sigma_{s_k}^2 {\bf u}_k^H {\bf u}_k} +
    \sum_{k=1}^K{\pmb\mu}_k^T(\Re\{{\bf H}'{\bf u}_k\} - \sqrt{\lambda_k}{\bf e}_k)
\end{align}
where ${\pmb\mu}_k$, $1\le k\le K$, are the nonnegative Lagrange multipliers.
Using Wirtinger calculus \cite{Brandwood83,GesbertTSP07,Koivunen10}
to take the derivative of the Lagrangian
(\ref{zfLagrangian_wl_eq}) with respect to the complex-valued precoding vectors
${\bf u}_k$, $1\le k\le K$, and writing the KKT conditions results in the
following equations for the stationary points of (\ref{zfBfModified_wl_eq}):
\begin{subequations}\label{zfKkt_wl_eq}
\begin{align}
    \label{zfKkt_wl_subeq1}
    &\sigma_{s_k}^2{\bf u}_k^* + \frac{1}{2}{\bf H}^{\prime T}{\pmb\mu}_k =
    {\bf 0}, \quad 1 \le k \le K \\
    \label{zfKkt_wl_subeq2}
    &\Re\{{\bf H}'{\bf u}_k\} - \sqrt{\lambda_k}{\bf e}_k = {\bf 0}, \quad 1 \le k \le K.
\end{align}
\end{subequations}
From (\ref{zfKkt_wl_subeq1}), and for $1 \le k \le K$
\begin{equation}\label{zfUIntermediate_wl_eq}
    {\bf u}_k = -\frac{1}{2\sigma_{s_k}^2} {\bf H}^{\prime H} {\pmb \mu}_k.
\end{equation}
Substituting (\ref{zfUIntermediate_wl_eq}) into (\ref{zfKkt_wl_subeq2}) yields
the corresponding Lagrange multipliers
\begin{equation}\label{zfLagrangeMultiplier_wl_eq}
    {\pmb \mu}_k = -2\sigma_{s_k}^2\sqrt{\lambda_k}
    [\Re\{{\bf H}'{\bf H}^{\prime H}\}]^{-1} {\bf e}_k.
\end{equation}
Using (\ref{zfLagrangeMultiplier_wl_eq}) in (\ref{zfUIntermediate_wl_eq})
and concatenating the obtained ZF precoding vectors to form a matrix, the widely
linear ZF precoder is obtained as
\begin{equation}\label{zfU_wl_eq}
    {\bf U}_{\text{ZF}} = {\bf H}^{\prime H} [\Re\{ {\bf H}' {\bf H}^{\prime H} \}]^{-1} {\pmb
    \Lambda}.
\end{equation}

Widely linear ZF precoding suffers from the lack of a constraint on the transmit power
at the expense of a fixed received power, which makes the total transmit power
depend on the channel characteristics.
To overcome this shortcoming of WL ZF, a simple heuristic approach is to introduce
a scaling factor $\gamma$ to normalize ${\bf U}$ and constrain
the total transmit power to $\tau$ \cite{Nossek05}.
In other words, setting
$\gamma ^2 \Tr({\bf U}_{\text{ZF}} {\bf R}_s {\bf U}_{\text{ZF}}^H) = \tau$,
which results in
\begin{equation}\label{zfScaling_wl_eq}
    {\gamma} = \sqrt{\frac{\tau} {\Tr({\bf H}^{\prime H} [\Re\{ {\bf H}' {\bf H}^{\prime H} \}]^{-1}
    {\pmb\Lambda}^2{\bf R}_s [\Re\{ {\bf H}' {\bf H}^{\prime H} \}]^{-1} {\bf H}' )}}  .
\end{equation}
Using the scaling factor (\ref{zfScaling_wl_eq}) to normalize ${\bf U}_{\text{ZF}}$
of (\ref{zfU_wl_eq}), results in the following normalized WL ZF precoding matrix
\begin{equation}
    {\bf U}_{\text{ZF}}^{\text{Norm}} = \frac{\sqrt{\tau}
    {\bf H}^{\prime H} [\Re\{ {\bf H}' {\bf H}^{\prime H} \}]^{-1}}
    {\!\sqrt{\!\Tr(\!{\bf H}^{\prime H} \![\Re\{ {\bf H}' {\bf H}^{\prime H} \}]^{\!-\!1}
    {\pmb\Lambda}^2{\bf R}_s [\Re\{ {\bf H}' {\bf H}^{\prime H} \}]^{\!-\!1} {\bf H}' \!)}}.
\end{equation}
It should be remarked that in general, using this approach or other power allocation
approaches such as water-filling results in diagonal matrix
$\Re\{{\bf W}{\bf H}{\bf U}\}$ not being necessarily
equal to ${\pmb \Lambda}$ as required by (\ref{wlZfConstraint_wl_eq}).

\subsection{WL Minimum Mean Square Error Precoding}\label{subSecMmse_wl}
In this section, we consider minimizing the sum of mean square errors (MSE) of users
by constraining the transmit power. Using the estimate ${\bf\hat s} = \Re\{\bf y\}$,
the sum MSE between the estimated signals and the desired signals can be written as
\begin{align}\label{mseNoTransformation_wl_eq}
    \nonumber
    &{\text{MSE}} = \E[\|{\bf\hat s}-{\bf s}\|_2^2]
    = \Tr(\Re\{{\bf H}'{\bf U}\} {\bf R}_s \Re\{{\bf U}^T{\bf H}^{\prime T}\}) \\
    &- 2 \Tr(\Re\{{\bf H}'{\bf U}\}{\bf R}_s) + \Tr(\frac{1}{2}{\bf R}_{z'} + {\bf R}_s).
\end{align}
Unfortunately, a similar Wirtinger calculus approach of Section \ref{subSecZf_wl} results
in ${\bf U}$ and ${\bf U}^*$ of WL MMSE being coupled in such a way that
a closed-form or a semi closed-form solution would not be possible.

To deal with the above widely linear minimum mean square error (MMSE) precoding
problem we employ the following two isomorphisms from the complex field to the real field:
\begin{subequations}\label{isomorphisms_wl_eq}
\begin{align}
    &{\bf U} \xrightarrow{\mathcal{T}_1}
    {\bf\bar U} = \left[\begin{array}{c}
    \Re\{{\bf U}\} \\ \Im\{{\bf U}\} \end{array} \right]  \\
    &{\bf H}' \xrightarrow{\mathcal{T}_2} {\bf\tilde H}' =
    \left[\begin{array}{cc}
    \Re\{{\bf H}'\} & -\Im\{{\bf H}'\} \end{array} \right].
\end{align}
\end{subequations}
Using these transformations 
the sum MSE in (\ref{mseNoTransformation_wl_eq}) can be equivalently expressed as
\begin{equation}\label{mse_wl_eq}
    {\text{MSE}} \!=\! \Tr({\bf\tilde H}'{\bf\bar U} {\bf R}_s
    {\bf\bar U}^T\!{\bf\tilde H}^{\prime T}\!) \!-\! 2
    \Tr({\bf\tilde H}'{\bf\bar U}{\bf R}_s\!) \!+\!
    \Tr(\!\frac{1}{2}{\bf R}_{z'} \!+\! {\bf R}_s\!).
\end{equation}
The total transmit power (\ref{txPower_wl_eq}) can also be rewritten as
\begin{equation}\label{txPowerTransformed_wl_eq}
    \E[\|{\bf x}\|^2_2] = \Tr({\bf\bar U}{\bf R}_s{\bf\bar U}^T).
\end{equation}

Having the MSE (\ref{mse_wl_eq}) and the transmit power (\ref{txPowerTransformed_wl_eq}),
in a fashion similar to the linear MMSE
precoder \cite{Nossek05}, the widely linear MMSE precoder
for the broadcast channel is obtained by solving
\begin{subequations}\label{wlMmseBf_wl_eq}
\begin{align}
    \label{wlMmseBf_wl_subeq1}
    &\min_{\bf\bar U} \text{MSE}\\
    \label{wlMmseBf_wl_subeq2}
    &{\text {subject to}} \quad \Tr({\bf\bar U}{\bf R}_s{\bf\bar U}^T) \le \tau.
\end{align}
\end{subequations}
It should be noted that constraint (\ref{wlMmseBf_wl_subeq2}) is not necessarily
active \cite{NocedalBook06}.

Ignoring the constant terms to save space, the Lagrangian function corresponding to the
optimization problem (\ref{wlMmseBf_wl_eq}) is
\begin{align}\label{mmseLagrangian_wl_eq}
    \nonumber
    &L({\bf\bar U}, \mu) = \Tr({\bf\tilde H}'{\bf\bar U} {\bf R}_s
    {\bf\bar U}^T{\bf\tilde H}^{\prime T}) - 2 \Tr({\bf\tilde H}'{\bf\bar U}{\bf R}_s)\\
    &+ \mu (\Tr({\bf\bar U} {\bf R}_s {\bf\bar U}^T) - \tau),
\end{align}
where $\mu$ is the non-negative Lagrange multiplier.
Therefore, the Lagrange dual function of (\ref{wlMmseBf_wl_eq}) is
$g(\mu) = \min_{{\bf\bar U}} L({\bf\bar U},\mu)$, and hence the corresponding dual problem is
\begin{align}
    \nonumber
    &\max_{\mu} g(\mu) \\
    &{\text {subject to}} \quad \mu \ge 0.
\end{align}

Inherently, the dual problem is a convex optimization problem with respect to
$\mu$. To solve the dual problem, similar to
\cite{BoydBook11,Yu06,Mokari10}, we take a dual ascent approach
which minimizes the Lagrangian (\ref{mmseLagrangian_wl_eq}) and maximizes
the dual function alternatingly.
By setting the derivative of the Lagrangian (\ref{mmseLagrangian_wl_eq}) with
respect to ${\bf\bar U}$ to zero, the minimizer could be obtained as
\begin{equation}\label{mmseU_wl_eq}
    {\bf\bar U}_{\text{MMSE\_Iter}} \!=\! ({\bf\tilde H}^{\prime T} {\bf\tilde H}' +
    \mu {\bf I}_{2M})^{-1} {\bf\tilde H}^{\prime T} \!=\!
    {\bf\tilde H}^{\prime T} ({\bf\tilde H}' {\bf\tilde H}^{\prime T} + \mu {\bf I}_{K})^{-1},
\end{equation}
where the second equality results from the matrix inversion lemma \cite{MoonBook99}.
From (\ref{mmseU_wl_eq}), it is obvious that $K\le 2M$, otherwise
the channels would not be linearly independent\footnote{
Linearly dependent channels indicate a degraded broadcast channel.}.
The dual ascent algorithm summarized in Table \ref{tableWLMMSE} is proposed to solve the WL MMSE
precoding problem.
To maximize the dual function, in each iteration of the proposed algorithm
the Lagrange multiplier
$\mu$ is updated in a way that it moves in the
direction of its steepest ascent, or derivative, as
\begin{align}\label{updateMuMmse_wl_eq}
    \mu^{l+1} = [\mu^l + \delta^l_{\mu}(\Tr({\bf\bar U} {\bf R}_s {\bf\bar U}^T)-\tau)]^+
\end{align}
where $l$ denotes the iteration number and $\delta^l_\mu$
indicates the sequence of positive scalar step sizes for $\mu$ \cite{Yu06,Mokari10}.
\begin{table}
    \centering
    \caption{Dual ascent algorithm for finding WL MMSE precoding matrix in broadcast channels}
    \label{tableWLMMSE}
    \begin{tabular}{l}
        \hline\hline
        \hspace{-.13in}
        \begin{minipage}[]{3.4in}
            \begin{algorithmic}
                \STATE
                \STATE {Initialize $\mu$.} \\
                \STATE $l = 1$.
                \REPEAT
                \STATE Compute ${\bf\bar U}$ using (\ref{mmseU_wl_eq}).
                \STATE Update $\mu$ using (\ref{updateMuMmse_wl_eq}).
                \STATE $l \gets l+1$.
                \UNTIL {${\bf\bar U}$ converges}
            \end{algorithmic}
        \end{minipage}
        \\
        \hline\hline
    \end{tabular}
\end{table}

It should be remarked that, besides the dual ascent approach, another
approach to address the WL MMSE precoding problem
is from the perspective of regularized zero-forcing \cite{Peel05}. This approach results
in the following WL MMSE precoding matrix:
\begin{equation}\label{mmseURegularized_wl_eq}
    {\bf\bar U}_{\text{MMSE}} =
    {\bf\tilde H}^{\prime T} ({\bf\tilde H}' {\bf\tilde H}^{\prime T} +
    \frac{1}{2\gamma} {\bf I}_{K})^{-1},
\end{equation}
where $\gamma = \frac{\tau}{K \sigma_z^2}$, assuming that
${\bf R}_z = \sigma_{z}^2 {\bf I}_K$.


\subsection{WL Maximum Signal to Leakage and Noise Ratio Precoding}\label{subSecMslnr_wl}
So far, we have developed widely linear MRT, ZF, and MMSE precoding. Similar to
linear MRT, ZF, and MMSE precoders (excluding the regularized zero-forcing approach) \cite{Nossek05},
it can be seen that their widely linear
counterparts also do not consider the effect of the receiver's additive noise
in calculating the precoding vectors. A conventional performance criterion in communications
systems which also reflects the effect of additive
noise is maximization of the signal to interference and noise ratio (SINR). However, finding
precoding vectors by maximizing SINR of each user is a prohibitively complex problem and
does not lead to a closed-form solution \cite{Shamai06,Schubert04}. 
On the other hand, signal to leakage and noise ratio (SLNR) is a relatively new metric, which
not only considers the effect of noise but also its maximization
results in a closed-form solution for the precoding vectors \cite{Sadek07WC}. 
In a broadcast channel with linear precoding, the power of the
leakage of user $k,~1 \le k \le K$, is
defined as the expected total power of the signal transmitted to user $k$ that
is leaked to other users' receivers, i.e.,
\begin{align}
    \nonumber
    &\text{Leakage}_k = \sum_{\substack{j=1 \\ j \ne k}}^K \E[|w_j {\bf h}_j {\bf u}_k s_k|^2] \\
    &= \sum_{\substack{j=1 \\ j \ne k}}^K \sigma_{s_k}^2 {\bf u}_k^H{\bf h}_j^{\prime H}
    {\bf h}'_j {\bf u}_k =  \sigma_{s_k}^2
    {\bf u}_k^H {\bf H}_{\bar k}^{\prime H} {\bf H}'_{\bar k} {\bf u}_k,
\end{align}
where
${\bf H}'_{\bar k} = [{\bf h}_1^{\prime H}, \cdots, {\bf h}_{k-1}^{\prime H},
{\bf h}_{k+1}^{\prime H}, \cdots, {\bf h}_K^{\prime H}]^H$ and ${\bf h}'_j$s were introduced
in Section \ref{subSecMrt_wl}.
Consequently, the SLNR of user $k$ is defined as the ratio between the expected power of the
desired part of the received signal
of that user and the combined expected noise and leakage powers \cite{Sadek07WC}:
\begin{equation}
    \frac{\sigma_{s_k}^2 {\bf u}_k^H {\bf h}_k^{\prime H} {\bf h}'_k {\bf u}_k}
    {\sigma_{s_k}^2 {\bf u}_k^H {\bf H}_{\bar k}^{\prime H} {\bf H}'_{\bar k} {\bf u}_k +
    \sigma_{z'_k}^2} .
\end{equation}

Considering one-dimensional modulation combined with widely linear processing, the
SLNR expression can be revised to accommodate only the effective part of the powers
on the estimation of the received signals:
\begin{equation}\label{slnrReal_wl_eq}
    \text{SLNR}_k = \frac{\sigma_{s_k}^2 (\Re\{{\bf h}'_k {\bf u}_k\})^2}
    {\sigma_{s_k}^2 \Re\{{\bf u}_k^H {\bf H}_{\bar k}^{\prime H}\}
    \Re\{{\bf H}'_{\bar k} {\bf u}_k\} +
    \frac{\sigma_{z'_k}^2}{2}} .
\end{equation}
To maximize the SLNR (\ref{slnrReal_wl_eq}), we again have to use the isomorphisms
in (\ref{isomorphisms_wl_eq}) to decouple ${\bf U}$ and ${\bf U}^*$ in the optimization problem.
Using these isomorphisms,
SLNR (\ref{slnrReal_wl_eq}) can be rewritten as
\begin{equation}\label{slnrTransformed_wl_eq}
    \text{SLNR}_k \!=\! \frac{\sigma_{s_k}^2 ({\bf\tilde h}'_k {\bf\bar u}_k)^2}
    {\sigma_{s_k}^2 \sum_{\substack{j=1\\j \ne k}}^K ({\bf\tilde h}'_j {\bf\bar u}_k)^2 \!+\!
    \frac{\sigma_{z'_k}^2}{2}} \!=\!
    \frac{\sigma_{s_k}^2 {\bf\bar u}_k^T {\bf\tilde h}_k^{\prime T} {\bf\tilde h}'_k {\bf\bar u}_k}
    {\sigma_{s_k}^2 {\bf\bar u}_k^T {\bf\tilde H}_{\bar k}^T {\bf\tilde H}_{\bar k} {\bf\bar u}_k
    \!+\! \frac{\sigma_{z'_k}^2}{2} } .
\end{equation}
Therefore, the WL maximum SLNR (MSLNR) precoding problem can be stated as
\begin{subequations}\label{mslnrBf_wl_eq}
\begin{align}
    &\max_{{\bf\bar u}_k} \text{SLNR}_k \\
    &{\text{subject to}} \quad \sigma_{s_k}^2 \|{\bf\bar u}_k\|_2^2 = \tau_k,
\end{align}
\end{subequations}
where $\text{SLNR}_k$ is given by (\ref{slnrTransformed_wl_eq}).
By casting (\ref{mslnrBf_wl_eq}) into a generalized Rayleigh quotient problem as
\begin{subequations}
\begin{align}
    &{\bf\bar u}_{k_\text{MSLNR}} = \argmax_{{\bf\bar u}_k}
    \frac{{\bf\bar u}_k^T {\bf\tilde h}_k^{\prime T} {\bf\tilde h}'_k {\bf\bar u}_k}
    {{\bf\bar u}_k^T ({\bf\tilde H}_{\bar k}^T {\bf\tilde H}_{\bar k}
    + \frac{\sigma_{z'_k}^2}{2\tau_k} {\bf I}_{2M}){\bf\bar u}_k } \\
    &{\text{subject to}} \quad \sigma_{s_k}^2 \|{\bf\bar u}_k\|_2^2 = \tau_k,
\end{align}
\end{subequations}
the solution to the WL MSLNR precoding problem is given by
\begin{equation}
    {\bf\bar u}_{k_\text{MSLNR}} = \frac{\sqrt{\tau_k}}{\sigma_{s_k}}
    {\bf v}_{\max}({\bf\tilde h}_k^{\prime T} {\bf\tilde h}'_k, {\bf Q}'_k),
\end{equation}
where ${\bf Q}'_k = {\bf\tilde H}_{\bar k}^{\prime T} {\bf\tilde H}'_{\bar k}
+ \frac{\sigma_{z'_k}^2}{2 \tau_k} {\bf I}_{2M}$
and ${\bf v}_{\max}({\bf\tilde h}_k^{\prime T} {\bf\tilde h}'_k, {\bf Q}'_k)$ is
the normalized eigenvector corresponding to the largest
generalized eigenvalue of ${\bf\tilde h}_k^{\prime T} {\bf\tilde h}'_k$ and ${\bf Q}'_k$.


\section{Semi-orthogonal User Selection for One-dimensional Modulation}\label{secSusom_downlink}
User selection is a complementary approach to transmit precoding to also deal with the
co-channel interference that arises due to spatial multiplexing.
When the total number of available users $K_T$
is large and the transmitter has data available for transmission to all $K_T$ users,
suppose that the transmitter selects $K < K_T$ users with sufficiently {\it good} channel conditions
for simultaneous transmission.
By selecting $K$ users with good channels, co-channel interference is
reduced, resulting in improved throughput and reliability.
In principle, the optimal user subset can be found by brute-force search over all
possible user subsets, although with prohibitive computational complexity.

Most existing multiuser linear precoding methods for MISO systems can only support
at most as many users as the number of transmit antennas $M$ \cite{Spencer04,Peel05,Nossek05}.
Consequently, existing user selection algorithms that tackle interference arising
in multiuser communications can only select at most $M$
users \cite{Goldsmith06,Evans06,Mao12}.
One such algorithm is semi-orthogonal user selection (SUS) \cite{Goldsmith06}.
The semi-orthogonal user selection algorithm
tries to select a user with a large channel
gain that is also nearly orthogonal to the channels of other selected users.
Ideally, all selected channels by SUS are orthogonal to one another and at the same time
have the largest gains among $K_T$ available channels. In the SUS algorithm, first the
user with the strongest channel among the available users is selected.
Then all the channels that are not nearly orthogonal to the previously selected channel
are removed from the set of available channels. This process is then repeated until either
the set of available channels is empty or until the number of selected channels is the same as
the number of transmit antennas.

As has been shown, for example in Section \ref{subSecMmse_wl},
WL precoding of 1D modulated signals is capable of
supporting more users than the number of transmit antennas. This motivates the
design of a user selection algorithm that can select more users than the number
of transmit antennas $M$.
To the best of our knowledge, the geometric user selection (GUS) algorithm proposed in
\cite{MajidArXiV16}
is the only existing user selection algorithm (based on the notion of interference
avoidance) that
can select more than $M$ users. Although the
computational complexity of the GUS algorithm is very low, it may not
be able to select a large enough number of users, particularly
when the total number of available users is sufficiently large.
This prompts us to devise a user selection algorithm with better performance.

Obviously, when the modulated signals are complex-valued, ${\bf h}_k$ and
${\bf h}_j$, $j\ne k$, are considered to be orthogonal if ${\bf h}_k{\bf h}_j^H = 0$, as is
assumed in SUS \cite{Goldsmith06}. However,
when the transmitted signals are one-dimensionally modulated on a real basis
function, the notion of orthogonality should be modified such that two channels
are considered to be orthogonal if \cite{ProakisBook01}:
\begin{equation}\label{semiOrthogonality_eq}
    \Re\{{\bf h}_k{\bf h}_j^H\} = 0.
\end{equation}
Therefore, SUS can be refashioned as in Table \ref{table_SUSOM} to incorporate
the above notion of orthogonality for one-dimensionally modulated signals.
\begin{table}[tp]
    \centering
    \caption{Semi-orthogonal user selection for one-dimensional modulation}
    \label{table_SUSOM}
    \begin{tabular}{l}
        \hline\hline
        \hspace{-.13in}
        \begin{minipage}[]{3.4in}
            \begin{algorithmic}
                \STATE
                \STATE {\bf \underline{Initialization:}} \\
                \STATE $i = 1$.
                \STATE $\mathcal{A} = \{1,\cdots, K_T\}$.
                \STATE $\mathcal{S} = \emptyset$.
                \STATE {\bf \underline{Main Body of Algorithm:}}\\
                \WHILE  {$i \le 2M$ and $\mathcal{A} \ne \emptyset$}
                    \STATE 1) $\pi_i = \operatornamewithlimits{argmax}\limits_{k
                    \in \mathcal{A}} \frac{\|{\bf \tilde h}_k\|_2}{\sigma_{z_k}}$ where
                    ${\bf \tilde h}_k$ is given by (\ref{projection_dl_eq}).\\
                    \STATE 2) $\mathcal{S} \gets \mathcal{S} \cup \{\pi_i\}.$
                    \STATE 3) ${\bf e}_{\pi_i} = {\bf \tilde h}_{\pi_i}$
                    \STATE 4) $\mathcal{A} \gets \mathcal{A} \setminus \{\pi_i\}$.
                    \STATE 5) $\mathcal{A} \gets \mathcal{A} \setminus
                    \{ \forall j \in \mathcal{A} |
                    \rm{dist}({\bf h}_j,{\bf e}_{\pi_i}) > \alpha\}$
                    \STATE 6) $i \gets  i+1$.
                \ENDWHILE
                \STATE \vspace{-.1in}
            \end{algorithmic}
        \end{minipage}
        \\
        \hline\hline
    \end{tabular}
\end{table}

We term the proposed algorithm in Table \ref{table_SUSOM} as semi-orthogonal user selection for
one-dimensional
modulation (SUSOM).
In SUSOM, at first, the set of available channels is initialized by all available
channels and the set of selected channels is set to be empty.
In Step 1, $\pi_i$ is set to be the index of the channel of user $k$
with the strongest effective
channel to noise ratio defined as $\frac{\|{\bf \tilde h}_k\|_2}{\sigma_{z_k}}$, where
$i$ is the iteration counter of the SUSOM algorithm.
The effective channel of user $k$, $\tilde {\bf h}_k$, is defined as the
component of ${\bf h}_k$ orthogonal{\footnote{Although we do not use this term, in
\cite{Troschke99}, this type of orthogonality as is defined in
(\ref{semiOrthogonality_eq}) is introduced as semi-orthogonality.}}
to the subspace spanned by the selected channels:
\begin{equation}\label{projection_dl_eq}
    {\bf \tilde h}_k \triangleq {\bf h}_k - \sum_{j=1}^{i-1}
    \frac{\Re\{ {\bf h}_k {\bf e}_{\pi_j}^H\}}{\|{\bf e}_{\pi_j}\|^2_2}
    {\bf e}_{\pi_j}.
\end{equation}
It should be remarked that in finding the component of a channel orthogonal to
the linear subspace spanned by the previously
selected channels, the projection of that channel on any element of that linear
subspace includes a real operator \cite{Troschke99}.
In Step 2, the channel with index $\pi_i$ is added to the
set of selected channels. In Step 3, the orthogonal component of the selected channel,
${\bf\tilde h}_{\pi_i}$, is
saved in ${\bf e}_{\pi_i}$, to be used later in Step 5 and in the following iterations of SUSOM
algorithm. In Step 4, the selected channel is removed from the set of available channels.
In Step 5, all the available channels in ${\mathcal{A}}$ that have distance
``$\rm{dist}$'' greater than
a predetermined threshold $\alpha$, $0 \le \alpha <1$, are removed from
the set of available channels
$\mathcal{A}$. The distance ``$\rm{dist}$'' which is defined as
\begin{equation}\label{distanceChannel_eq}
    \rm{dist}({\bf h}_j,{\bf e}_{\pi_i}) \triangleq
    \frac{|\Re\{ {\bf h}_j {\bf e}_{\pi_i}^H\}|}{\|{\bf h}_j\|_2 \|{\bf e}_{\pi_i}\|_2},
\end{equation}
measures the orthogonality of ${\bf h}_j$ and ${\bf e}_{\pi_i}$.
In other words, Step 5 results in semi-orthogonality of the selected channels.
It should be remarked that if the real operator in the definition (\ref{distanceChannel_eq})
is absent, it indicates the cosine of the principal angles between ${\bf h}_j$ and
${\bf e}_{\pi_i}$ \cite{Barg02}. Ideally, $\rm{dist}({\bf h}_j,{\bf e}_{\pi_i})$ should
be zero, i.e., ${\bf h}_j$ and ${\bf e}_{\pi_i}$ should be orthogonal to each other.

In SUSOM, since the notion of orthogonality is relaxed to only consider the real part,
it is expected that the number of selected users can be greater than $M$. Therefore, we
have the following claim:
\begin{claim}\label{claimMaximumUsers_wl}
    If the orthogonality is defined as (\ref{semiOrthogonality_eq}), the maximum
    number of channels that are mutually orthogonal is $2M$.
\end{claim}
\begin{IEEEproof}
    See Appendix \ref{subSecAppendClaimMaximumUsers_wl}.
\end{IEEEproof}

\begin{remark}
    It should be remarked that the complexity order of SUSOM is $K_TM^3$, i.e.,
    the same as that of the SUS algorithm \cite{Mao12}.
\end{remark}

\section{Numerical Results}\label{secResults_wl}
\subsection{WL Precoding}
We consider a multiple-input single-output broadcast channel (BC)
with a 4-antenna transmitter and four single-antenna users.
The transmitter is assumed to send independent 4-PAM signals to the users
simultaneously and at the same carrier frequency.
The channel gains are assumed to be quasi static and follow a
Rayleigh distribution with unit variance. In other words, each element of the channel
is generated as a zero-mean and unit-variance i.i.d.\ CSCG random variable.
Since our focus is on various transmit precoding methods rather
than on the effects of channel estimation, we assume that perfect CSI of all channels is
available at the transmitter \cite{Peel05,Sadek07WC}.
At the receiver, an i.i.d.\ Gaussian noise is added to the received signal.
All simulations are performed over 10,000 different channel realizations and at
each channel realization a block of 1,000 symbols is transmitted to each user.
The above set up is used for all of the following simulations unless indicated otherwise.

\begin{figure}[tp]
  \centering
  \includegraphics[width=3.49in]{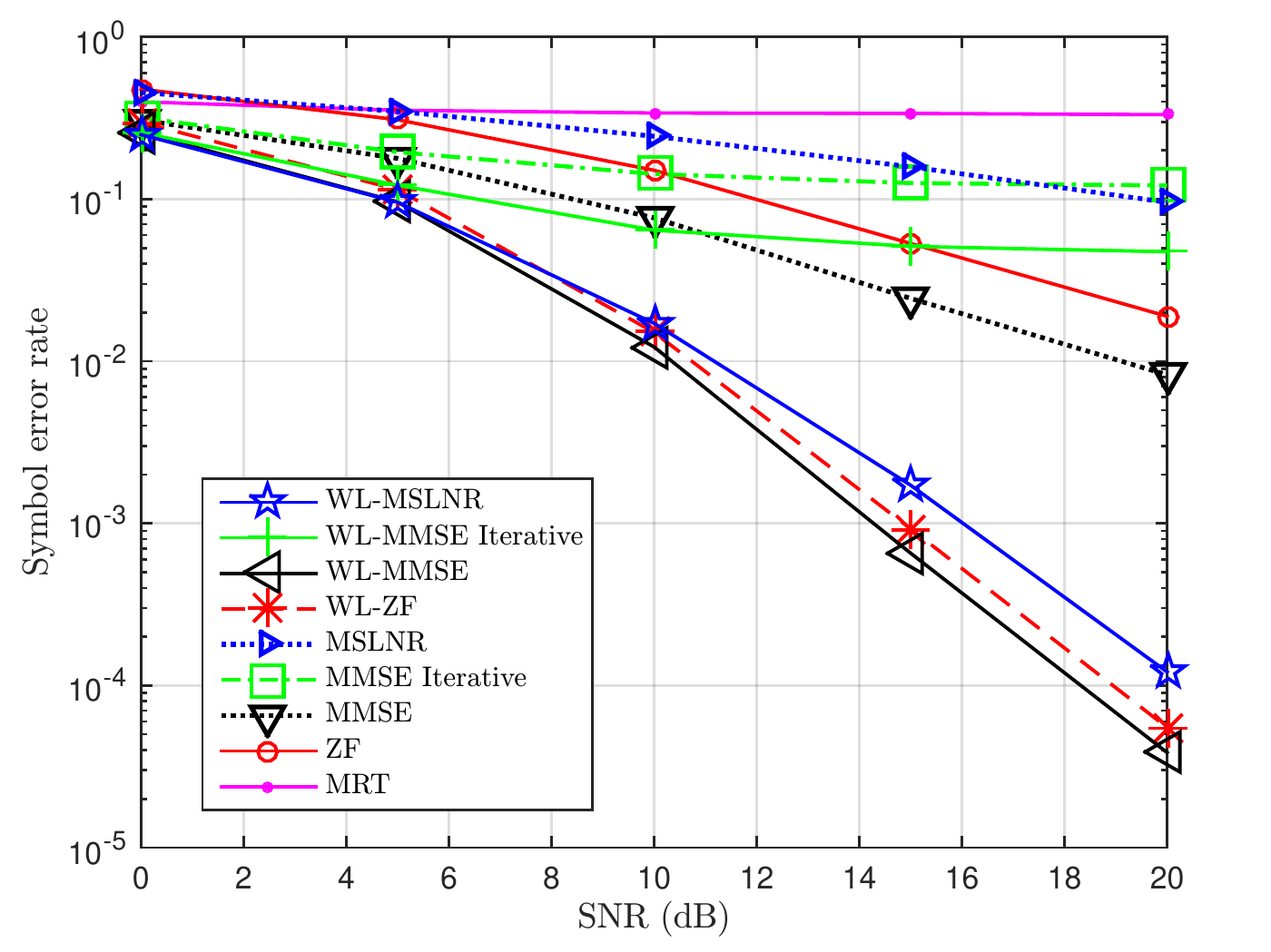}\\ 
  \caption{Average symbol error rates of users for $M=4$ transmit antennas and $K=4$ users
  with 4-PAM modulation. The MRT, ZF, and MMSE iterative precoding methods
  are given \cite{Nossek05}, MMSE precoding is given in \cite{Peel05}, and
  MSLNR precoding is given in \cite{Sadek07WC}.}
  \label{fig_wlPe}
\end{figure}
Fig. \ref{fig_wlPe} compares the average symbol error rates of linear
MRT, ZF, MMSE, iterative MMSE, and MSLNR precoding and their widely
linear counterparts.
As is expected, all the proposed widely linear precoding methods substantially
outperform their linear counterparts.
Moreover, it can be seen that the best performances
are achieved by WL MMSE, WL ZF, and WL MSLNR. Maximum ratio transmission,
which can be considered as both a linear and a widely linear processing
technique, does not exhibit good performance in high SNRs,
as expected. Similar to iterative MMSE \cite{Nossek05}, as SNR increases,
iterative WL MMSE also reaches an error floor very soon and does not show a promising performance.
Compared to MMSE, WL MMSE shows a gain
of about 9.2 dB at the error probability of $8.25 \times 10^{-3}$, which
demonstrates substantial performance improvement of WL processing compared to that of
linear processing.
It is interesting to note that the high SNR and low SNR performances of the
investigated linear precoding methods differ.
At higher SNRs, the best performance is achieved from top to bottom
by MMSE, ZF, MSLNR, iterative MMSE, and MRT.
These results are consistent with the results obtained in \cite{Nossek05}.
In a similar fashion, if widely linear methods are compared to one another, at higher SNRs
from best to worst the order of performance is WL MMSE, WL ZF,
WL MSLNR, iterative WL MMSE, and MRT.
An interesting remark on these comparisons is that widely linear processing
substantially benefits MSLNR precoding.
Linear MSLNR precoding, which in the above simulation setting did not perform well,
significantly improved by using widely linear processing.

\begin{figure}[tp]
  \centering
  \includegraphics[width=3.49in]
  {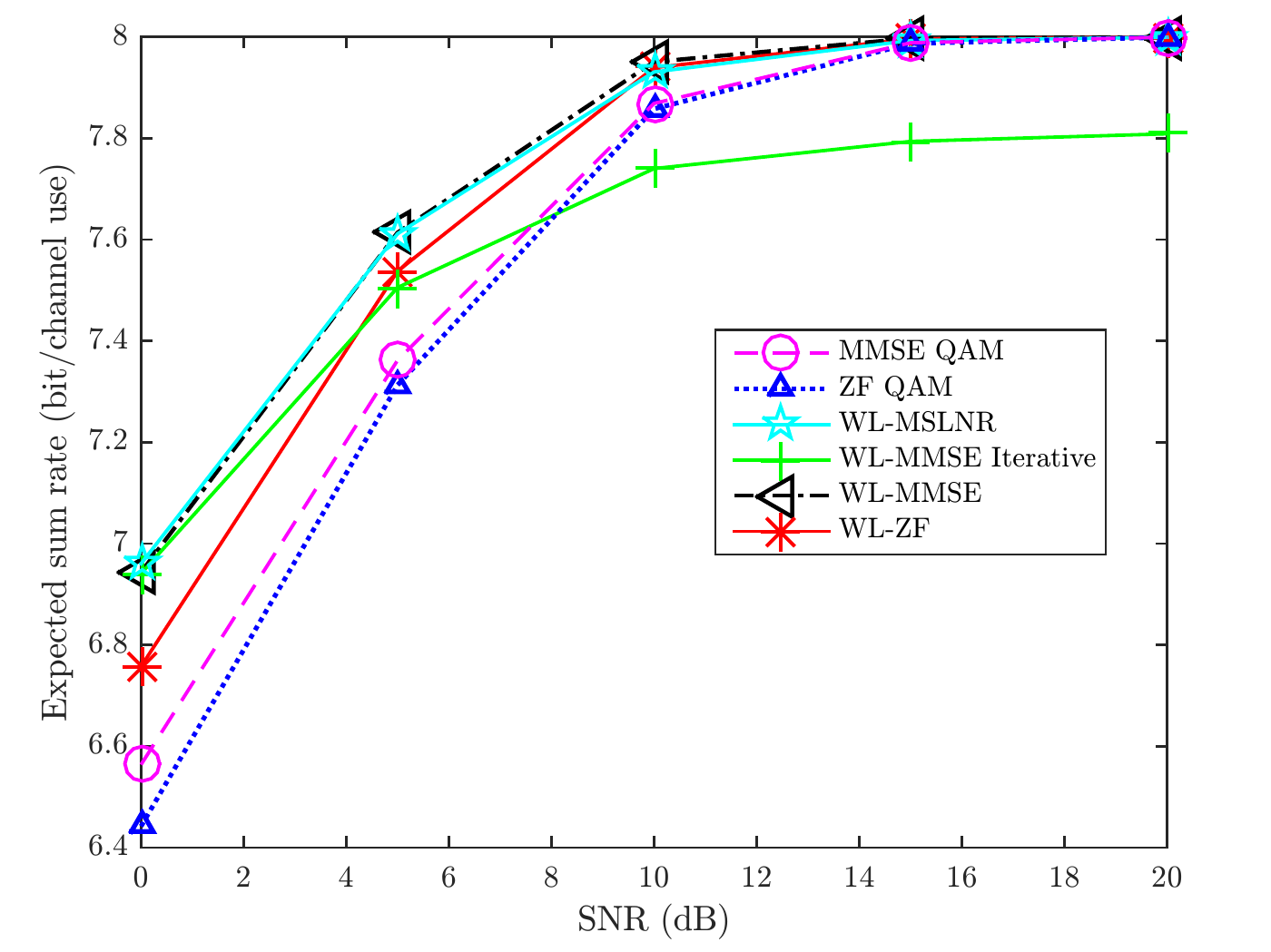}\\ 
  \caption{Average sum rates when $M=4$ and $K=2$ with 16-QAM modulated signals
  for ZF-QAM and MMSE-QAM and when $M=4$ and $K=4$ with 4-PAM modulated signals for
  the simulated widely linear precoding methods.}
  \label{fig_wlRateBit}
\end{figure}
In Fig. \ref{fig_wlPe}, it was shown that widely linear precoding of
one-dimensionally modulated signals outperforms linear precoding of
one-dimensionally modulated signals.
It would also be instructive to compare the performance of
widely linear precoding of one-dimensionally modulated signals
with that of linear precoding of two-dimensionally modulated signals.
We use system throughput for this comparison.
Fig. \ref{fig_wlRateBit} depicts the expected sum rates of four users
with 4-PAM modulation employing the proposed widely linear precoding methods and
the sum rate of two users with 16-QAM modulation employing linear precoding methods.
Theoretically, four users with 4-PAM modulation and
two users with 16-QAM modulation, both achieve a maximum sum rate of
8 bits/channel use. Therefore,
it is very interesting to observe that WL ZF and WL MMSE precoding of four
4-PAM modulated users outperform ZF and MMSE precoding of two 16-QAM modulated
users. Widely linear MSLNR precoding also outperforms
both ZF and MMSE precoding, at all simulated SNRs. The fact that iterative WL MMSE
seems to be unable to achieve the sum rate of 8 bits/channel use
is also consistent with our findings in Fig. \ref{fig_wlPe} which
exhibits the error floor of iterative WL MMSE precoding.

\begin{figure}[tp]
  \centering
  \includegraphics[width=3.49in]
  {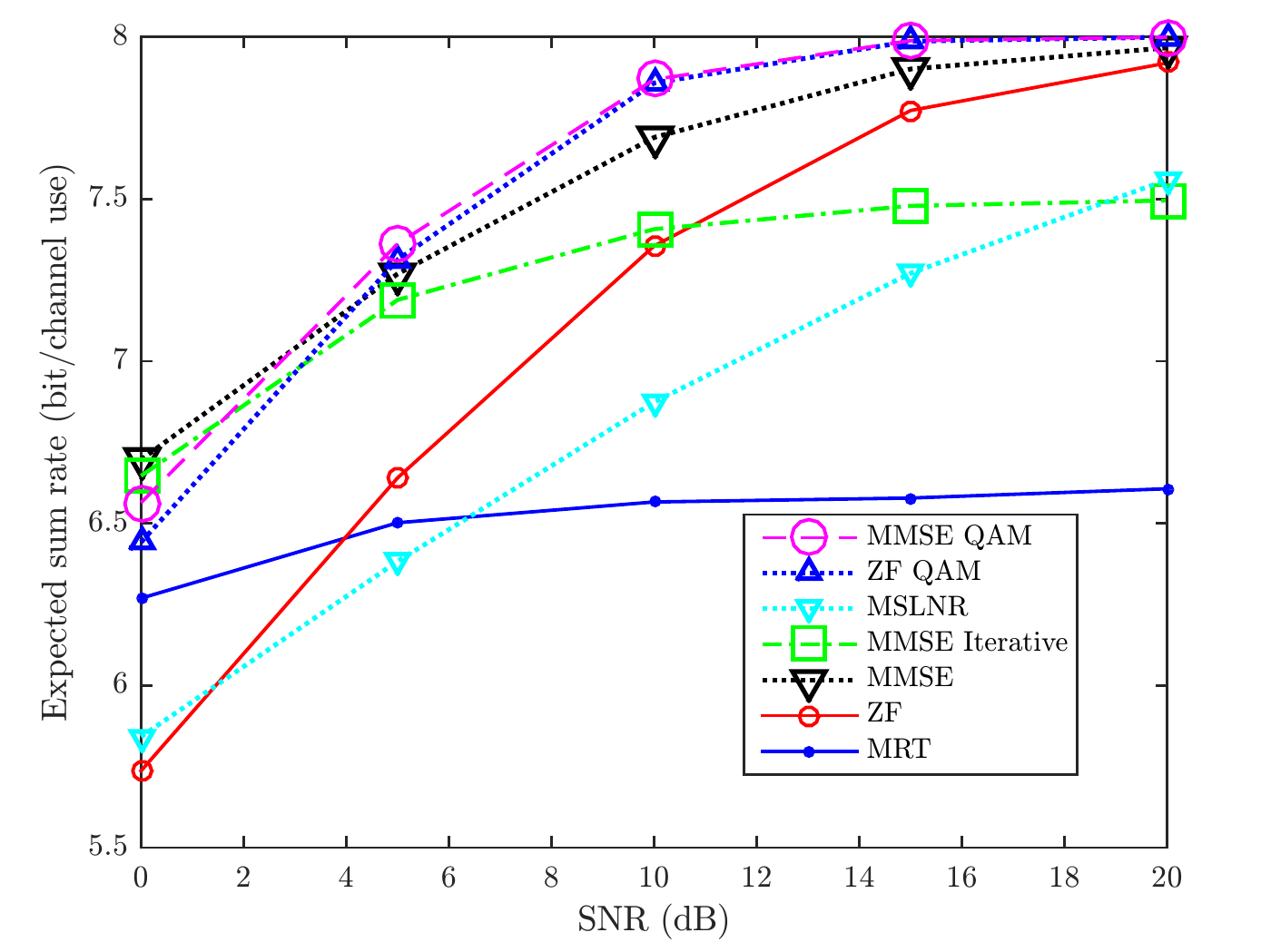}\\ 
  \caption{Average sum rates when $M=4$ and $K=2$ with 16-QAM modulated signals
  for ZF-QAM and MMSE-QAM and when $M=4$ and $K=4$ with 4-PAM modulated signals for
  the other simulated linear precoding methods.}
  \label{fig_wl_linearRateBit}
\end{figure}
To provide a more complete set of comparisons,
we also present Fig. \ref{fig_wl_linearRateBit} which depicts the expected
sum rates of linear precoding of four users with 4PAM modulation and
ZF and MMSE precoding of two users with 16-QAM modulation. At high SNRs,
the expected sum rates achieved by MMSE and ZF precoding of two 16-QAM
modulated users is higher than any other combination of modulation and
precoding, while at low SNRs, the expected sum rate of MMSE
precoding of four 4-PAM modulated users is higher than any
other combination of modulation and precoding. As expected from Fig.
\ref{fig_wlPe}, MRT, iterative MMSE, and MSLNR precoding methods
do not perform as well as other precoding methods. By comparing
Figs. \ref{fig_wlRateBit} and \ref{fig_wl_linearRateBit}, it
becomes clear that, at all simulated SNRs, all the proposed widely linear precoding methods
achieve higher bit rates compared to their linear counterparts.
This result is compatible with our findings in Fig. \ref{fig_wlPe}.

\subsection{User Selection}
\begin{figure}[tp]
  \centering
  \includegraphics[width=3.49in]{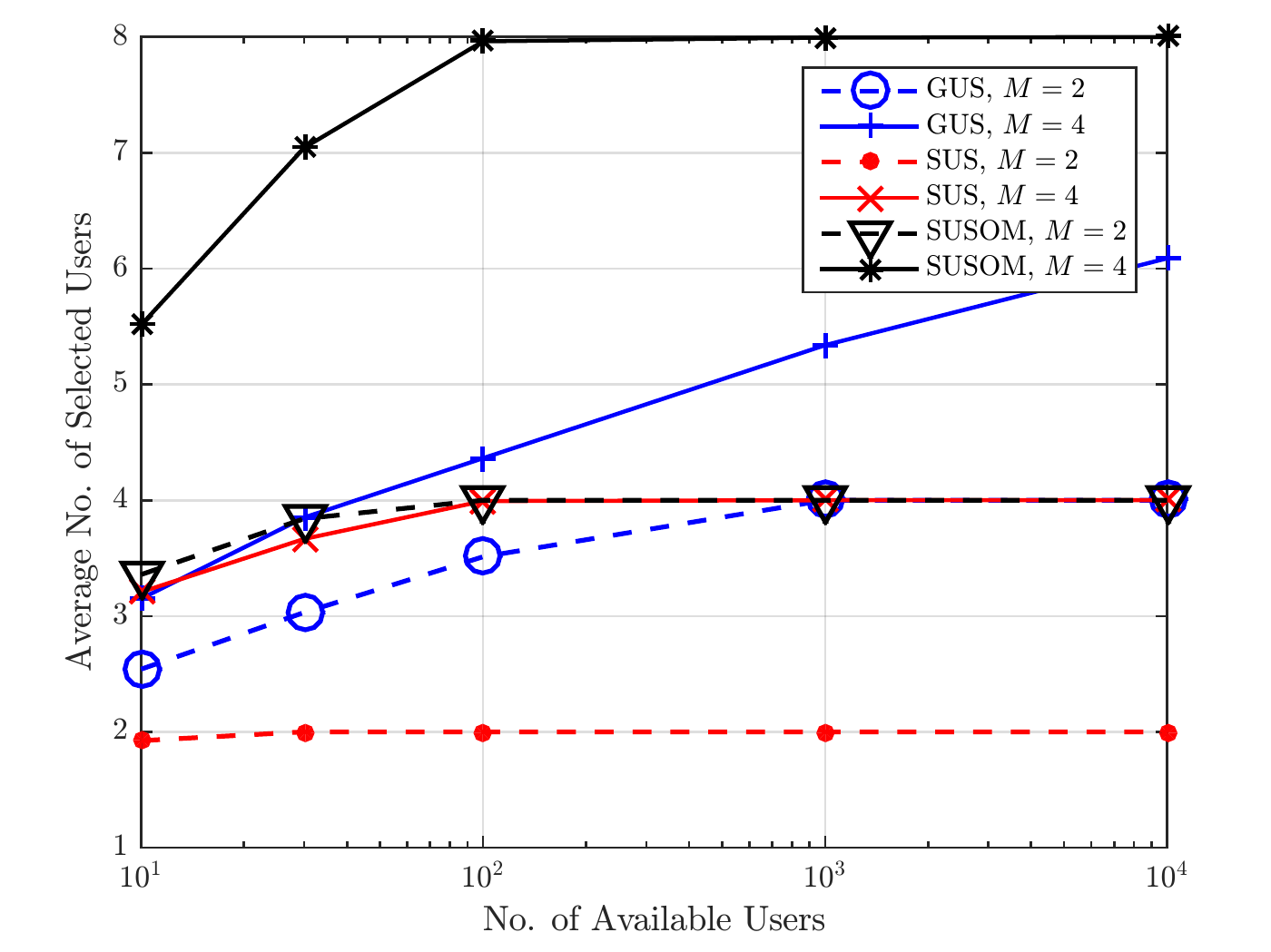}\\
  \caption{Average numbers of selected users with SUS \cite{Goldsmith06}, GUS \cite{MajidArXiV16},
  and SUSOM vs. total number of available users for $M = 2,~4$ transmit antennas.}
  \label{fig_wlSusomNoSelectedUsers}
\end{figure}
In this section, we evaluate the performance of the proposed
SUSOM algorithm of Table \ref{table_SUSOM}. In Fig. \ref{fig_wlSusomNoSelectedUsers},
the performance of SUSOM is compared to that of the SUS algorithm of \cite{Goldsmith06}
and the GUS algorithm of \cite{MajidArXiV16}, 
for $M = 2, 4$ transmit antennas when one
receive antenna is employed at each user. Each curve is averaged over 1,000 different
channel realizations. It can be observed in Fig. \ref{fig_wlSusomNoSelectedUsers} that as the
number of available users increases, the number of selected
users for all three algorithms increases until saturation.
As can be seen, the SUS algorithm could select at most $M$ users for both cases of
$M=2,~3$ antennas when $K$ is large enough.
On the other hand, the proposed SUSOM algorithm can select up to at most $2M$ users, i.e.,
twice of that of the SUS algorithm.
The GUS algorithm is also expected to saturate at $2M$ users, as it does
for $M=2$ transmit antenna scenario.
Nevertheless, it does not reach saturation in the scenario with $M=4$ transmit
antennas even with up to 10,000 available users{\footnote{While it is not practical to service
10,000 users with one transmitter, this large number is only for
illustration purposes to gain insight into the system.}}. This is an indication of the
slow saturation rate of GUS with respect to the number of available users,
which in turn indicates that
GUS algorithm does not always find the best set of users compared to SUSOM,
despite the fact that it can select more users than SUS.
It is interesting to observe that even before saturation SUSOM outperforms
both GUS and SUS.
For example, when $M=4$ and there are only 10 users available,
the average number of selected users is 5.51 with SUSOM, 3.15 with GUS, and 3.2
with SUS.

\begin{figure}[tp]
  \centering
  \includegraphics[width=3.49in]{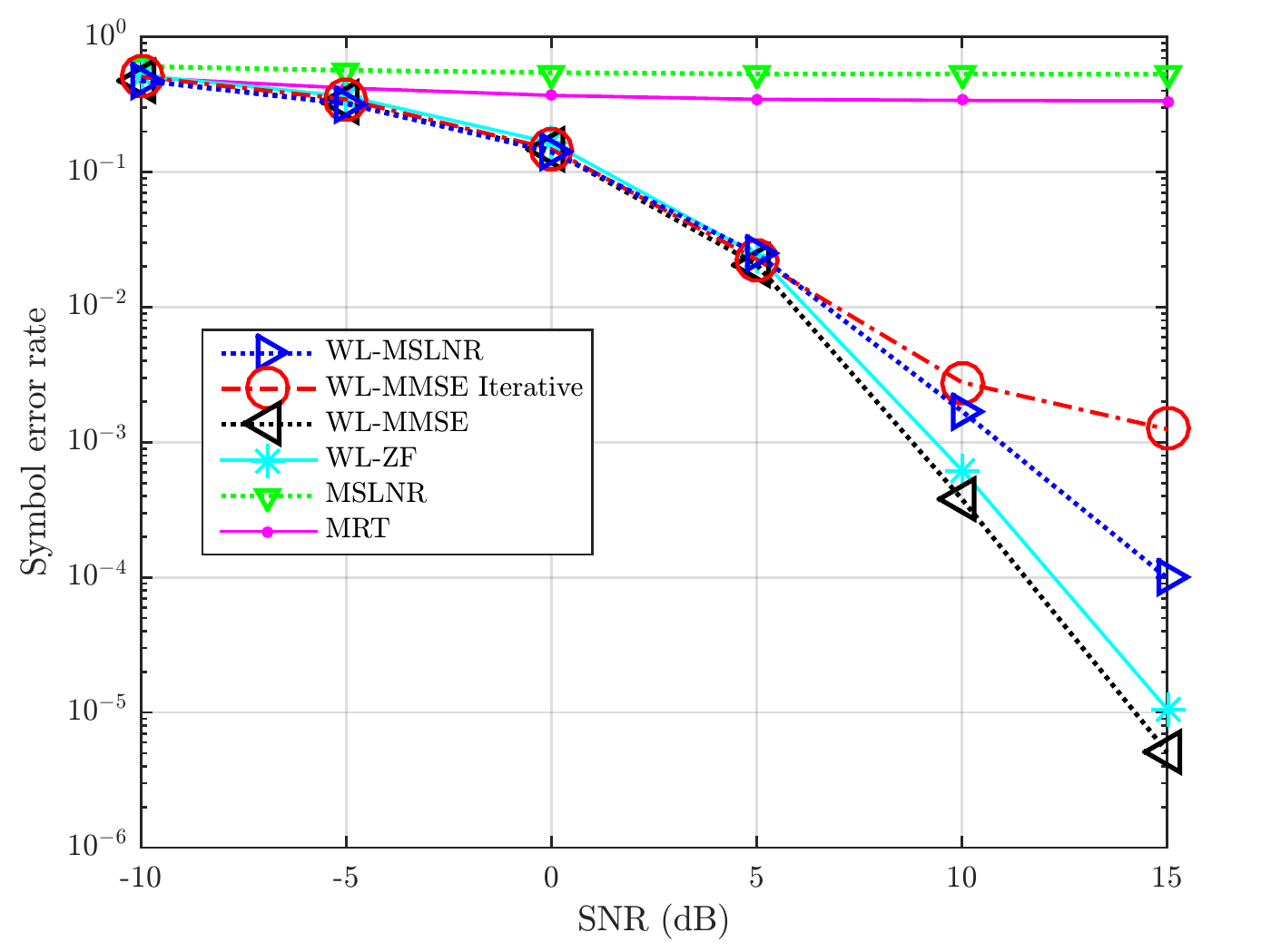}\\
  \caption{Average symbol error rates of users for $M=4$ and
  $K_T=100$ available users assuming the SUSOM algorithm is employed and all
  users transmit 4-PAM signals.}
  \label{fig_wlSusomPe}
\end{figure}
Fig. \ref{fig_wlSusomPe} compares the average symbol error rates of
MRT, MSLNR, WL ZF, WL MMSE, iterative WL MMSE, and WL MSLNR precoding
when the SUSOM algorithm is employed.
It is assumed that at first the SUSOM algorithm selects a set of users out of
$K_T=100$ available users and then the transmitter broadcasts information
to the selected users using the above precoding methods.
From Fig. \ref{fig_wlSusomNoSelectedUsers} it is known that using SUSOM algorithm the
average number of selected users is 7.96 when $K_T=100$, i.e., more than the number
of transmit antennas (system is overloaded).
Therefore, we do not perform linear ZF, MMSE, and iterative MMSE
for these cases, since they are not designed to work on overloaded
systems in their presented form.
We also do not show the error probabilities
of the above precoding methods combined with GUS and SUS algorithms, since
GUS and SUS algorithms select different numbers of users compared to SUSOM,
and therefore comparing the error probabilities in such a case would not bear
a meaningful interpretation. However, if we
compare Fig. \ref{fig_wlSusomPe} and Fig. \ref{fig_wlPe}, it can be observed
that the SUSOM algorithm not only selects more users than
that in Fig. \ref{fig_wlPe}, but also all the investigated precoding methods
under SUSOM achieve a better symbol error rate compared to those of Fig. \ref{fig_wlPe}.

\begin{figure}[tp]
  \centering
  \includegraphics[width=3.49in]
  {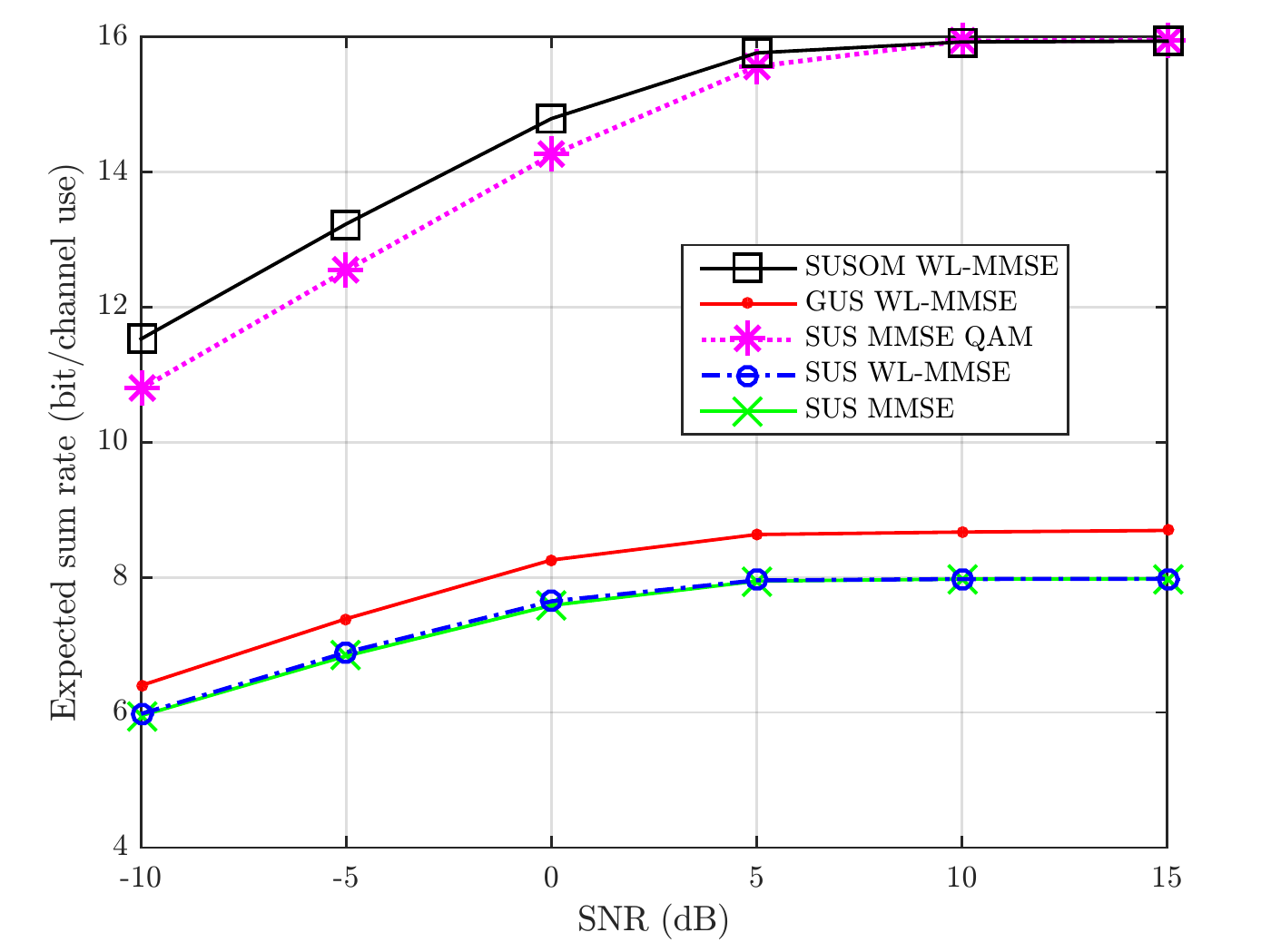}\\
  \caption{Average sum rates when $M=4$ and $K_T=100$ with 16-QAM modulated signals
  for SUS MMSE QAM and when $M=4$ and $K_T=100$ with 4-PAM modulated signals for
  the other user selection and precoding methods.
  The the MMSE precoding is given in \cite{Peel05},
  the SUS algorithm is given in \cite{Goldsmith06}, and the GUS algorithm is given in \cite{MajidArXiV16}.}
  \label{fig_wlSusomRateBit}
\end{figure}
As observed in Fig. \ref{fig_wlSusomPe}, since the error probability
alone is not a good indicator of the performance
when there are different numbers of users in the system, Fig. \ref{fig_wlSusomRateBit}
is provided to gain more insight into the relative
performances of the user selection algorithms.
In Fig. \ref{fig_wlSusomRateBit}, the expected sum rates of 4-PAM modulated
users for combinations of SUS with MMSE, SUS with WL MMSE, GUS with WL MMSE, and SUSOM with
WL MMSE are presented. In addition, the expected sum rate of 16-QAM modulated users
with SUS algorithm and MMSE precoding
is also plotted for comparison.
As can be seen, at all simulated SNRs, the combination of SUSOM with 4-PAM and WL MMSE
achieves the highest throughput.
It can be seen in Fig. \ref{fig_wlSusomRateBit} that
as SNR increases, the achievable expected throughput approaches limits
determined by the average numbers of selected users
and the order of modulation. For this example,
at higher SNRs the achievable throughput of both SUSOM with 4-PAM and SUS with 16-QAM
are 16 bits/channel use.

\section{Conclusion}\label{secConclusion_wl}
In this paper, we proposed a widely linear (WL) transmit precoding design for
one-dimensionally
modulated signals in a standard broadcast communications channel.
Closed-form solutions for
the precoders of the WL MRT and the WL ZF were obtained by using complex-domain
analysis and closed-form solutions of WL MMSE and WL MSLNR were
obtained by analysis of the composite real representation.
It was shown that WL ZF and WL MMMSE precoders can properly operate even if the
number of users is twice as large as the number of transmit antennas,
as opposed to linear ZF and MMSE precoders which can only support as many users as
the number of transmit antennas. We also developed a user selection algorithm,
compatible with widely linear precoding,
that can select twice as many users as the number of transmit antennas.
It has been shown that WL precoding outperforms linear precoding. Moreover, it has been shown that
widely linear precoding in conjunction with the proposed semi-orthogonal user
selection algorithm for one-dimensional modulation (SUSOM) also outperforms
linear precoding in conjunction with semi-orthogonal user selection
algorithm (SUS).


\appendix[Proof of Claim \ref{claimMaximumUsers_wl}]
\label{subSecAppendClaimMaximumUsers_wl}
Let us consider the following isomorphism from the complex field to the real field:
\begin{align}
    \nonumber
    &{\bf h}_k \in \mathbb{C}^{1\times M} \xrightarrow{\mathcal{T}_1}
    {\bf\bar h}_k = \left[\begin{array}{cc}
    \Re\{{\bf h}_k\} & \Im\{{\bf h}_k\} \end{array} \right] \in
    \mathbb{R}^{1\times 2M}, \\
    & 1 \le k \le K.
\end{align}
Then, it is obvious that
\begin{equation}
\Re\{{\bf h}_k{\bf h}_j^H\} = {\bf\bar h}_k {\bf\bar h}_j^T.
\end{equation}
If we define the $K \times 2M$ matrix  ${\bf\bar H}$ as
\begin{equation}
    {\bf\bar H} = \left[\begin{array}{c} {\bf\bar h}_1 \\ \vdots \\ {\bf\bar h}_K
    \end{array}  \right],
\end{equation}
then we have the following lemma:
\begin{lemma}
    All $K$ channels are mutually orthogonal if and only if the $K \times K$ matrix
    ${\bf\bar H}{\bf\bar H}^T$ is a full rank diagonal matrix.
\end{lemma}
\begin{IEEEproof}
    If ${\bf\bar H}{\bf\bar H}^T$ is a full rank diagonal matrix, then it is equivalently
    represented by $\diag(\allowbreak \|{\bf\bar h}_1\|_2^2, \ldots, \|{\bf\bar h}_K\|_2^2)$.
    In other words, ${\bf\bar h}_k {\bf\bar h}_j^T = 0$, $1 \le j \ne k \le K$,
    i.e., the channels are mutually
    orthogonal. On the other hand, $\|{\bf\bar h}_k\|_2 \ne 0$ with probability one and
    if the channels are mutually orthogonal then ${\bf\bar h}_k {\bf\bar h}_j^T = 0$,
    $1 \le j \ne k \le K$, which results in
    ${\bf\bar H}{\bf\bar H}^T$ being a full rank diagonal matrix.
\end{IEEEproof}
\noindent On the other hand, it is known that $\rank ({\bf\bar H}{\bf\bar H}^T) =
\rank({\bf\bar H}) \le \min(K,2M) \le 2M$. Therefore, the maximum number of
mutually orthogonal channels is $K=2M$.



\end{document}